\def\mode{1}  
\if 0\mode
  \documentclass[sigconf]{acmart}
  
  \definecolor{revisionColor}{rgb}{1, 0.45, 0.0}
  \definecolor{removedColor}{rgb}{0.85, 0.36, 0.22}
  \newcommand{\revision}[1]{{\color{revisionColor}#1}}
  \newcommand{\removed}[1]{{\color{removedColor}\sout{#1}}}
\else
  \documentclass[sigconf]{acmart}
  \newcommand{\revision}[1]{{\color{black}#1}}
  \newcommand{\removed}[1]{}
\fi

\usepackage[normalem]{ulem}
\usepackage{color}
\usepackage{float}
\usepackage{subcaption}
\usepackage{xspace} 
\usepackage{multirow}
\usepackage{array}
\usepackage{enumitem}
\usepackage{wrapfig}

\definecolor{burntorange}{rgb}{0.8, 0.33, 0.0}
\definecolor{byzantium}{rgb}{0.44, 0.16, 0.39}
\definecolor{byzantine}{rgb}{0.74, 0.2, 0.64}

\AtBeginDocument{%
  }

\copyrightyear{2025}
\acmYear{2025}
\setcopyright{cc}
\setcctype{by}
\acmConference[IUI '25]{30th International Conference on Intelligent User Interfaces}{March 24--27, 2025}{Cagliari, Italy}
\acmBooktitle{30th International Conference on Intelligent User Interfaces (IUI '25), March 24--27, 2025, Cagliari, Italy}
\acmDOI{10.1145/3708359.3712085}
\acmISBN{979-8-4007-1306-4/25/03}

\newcommand{\systemname}{Gensors\xspace} 
\newcommand{\exampleScenarioUser}{Emma\xspace} 

\newcommand{\fig}{Fig.\xspace}

\newcommand{\userquote}[1]{``\textit{#1}''}
\newcommand{\promptText}[1]{{\small\texttt{#1}}}

\begin{document}


\title{\systemname: Authoring Personalized Visual Sensors with Multimodal Foundation Models and Reasoning}




\author{Michael Xieyang Liu}
\authornote{Equal contribution.}
\affiliation{%
  \institution{Google DeepMind}
  \city{Pittsburgh, PA}
  \country{USA}
  }
\email{lxieyang@google.com}

\author{Savvas Petridis}
\authornotemark[1]
\affiliation{%
  \institution{Google DeepMind}
  \city{New York, NY}
  \country{USA}
  }
\email{petridis@google.com}

\author{Vivian Tsai}
\affiliation{%
  \institution{Google DeepMind}
  \city{Mountain View, CA}
  \country{USA}
  }
\email{vivtsai@google.com}

\author{Alexander J. Fiannaca}
\affiliation{%
  \institution{Google DeepMind}
  \city{Seattle, WA}
  \country{USA}
  }
\email{afiannaca@google.com}

\author{Alex Olwal}
\affiliation{%
  \institution{Google Research}
  \city{Mountain View, CA}
  \country{USA}
  }
\email{olwal@acm.org}

\author{Michael Terry}
\affiliation{%
  \institution{Google DeepMind}
  \city{Cambridge, MA}
  \country{USA}
  }
\email{michaelterry@google.com}

\author{Carrie J. Cai}
\affiliation{%
  \institution{Google DeepMind}
  \city{Mountain View, CA}
  \country{USA}
  }
\email{cjcai@google.com}

\renewcommand{\shortauthors}{Liu et al.}

\begin{abstract}
Multimodal large language models (MLLMs), with their expansive world knowledge and reasoning capabilities, present a unique opportunity for end-users to create \revision{personalized AI} sensors capable of reasoning about complex situations. A user could describe a desired sensing task in natural language (e.g., ``let me know if my toddler is getting into mischief in the living room''), with the MLLM analyzing the camera feed and \revision{responding within just seconds}. 
In a formative study, we found that users saw substantial value in defining their own sensors, yet struggled to articulate their unique personal requirements to the model and debug the sensors through prompting alone. 
To address these challenges, we developed \systemname, a system that empowers users to define \revision{customized} sensors supported by the reasoning capabilities of MLLMs. \systemname 1) \revision{assists users in eliciting requirements} through both automatically-generated and manually created sensor criteria, 2) \revision{facilitates} debugging by allowing users to isolate and test individual criteria in parallel, 3) suggests additional criteria based on \revision{user-provided images}, and 4) \revision{proposes} test cases to help users ``stress test'' sensors on potentially unforeseen scenarios. 
In a 12-participant user study, \revision{users reported significantly greater sense of control, understanding, and ease of communication} when defining sensors using \systemname. 
Beyond addressing model limitations, \systemname supported users in debugging, eliciting requirements, and expressing unique personal requirements to the sensor through criteria-based reasoning; 
it also helped uncover users' own ``blind spots'' by \revision{exposing overlooked criteria and revealing unanticipated failure modes}. 
Finally, we describe insights into how unique characteristics of MLLMs–such as hallucinations and inconsistent responses–can \revision{impact} the sensor-creation process. 
Together, these findings contribute to the design of future MLLM-powered sensing systems that are intuitive and customizable by everyday users.\looseness=-1

\end{abstract}





\begin{teaserfigure}
\centering
\includegraphics[width=1\textwidth]{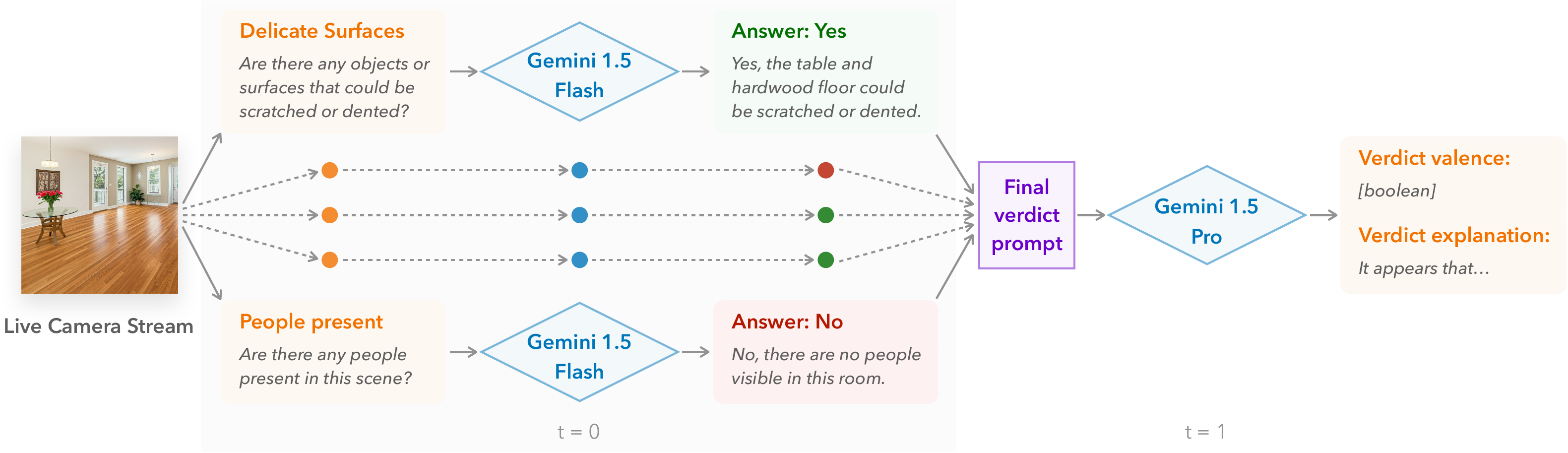}
\vspace{-7mm}
\caption{\systemname introduces a novel workflow to define and refine criteria for a sensor to monitor a live camera stream, analyze it using an MLLM, and provide a final verdict based on user-configurable logic and examples. The system uses a two-stage pipeline, where Gemini 1.5 Flash (optimized for speed) is called for all criteria to produce a collection of answers (t=0), which are subsequently sent to Gemini 1.5 Pro (optimized for reasoning capabilities and support for large context window), which is prompted to provide a final verdict (t=1). 
}
\label{fig:system}
\end{teaserfigure} 

\maketitle

%
%
\section{Introduction}\label{sec:intro}

\begin{figure*}[t]
\vspace{-1mm}
\centering
\includegraphics[width=1.0\textwidth]{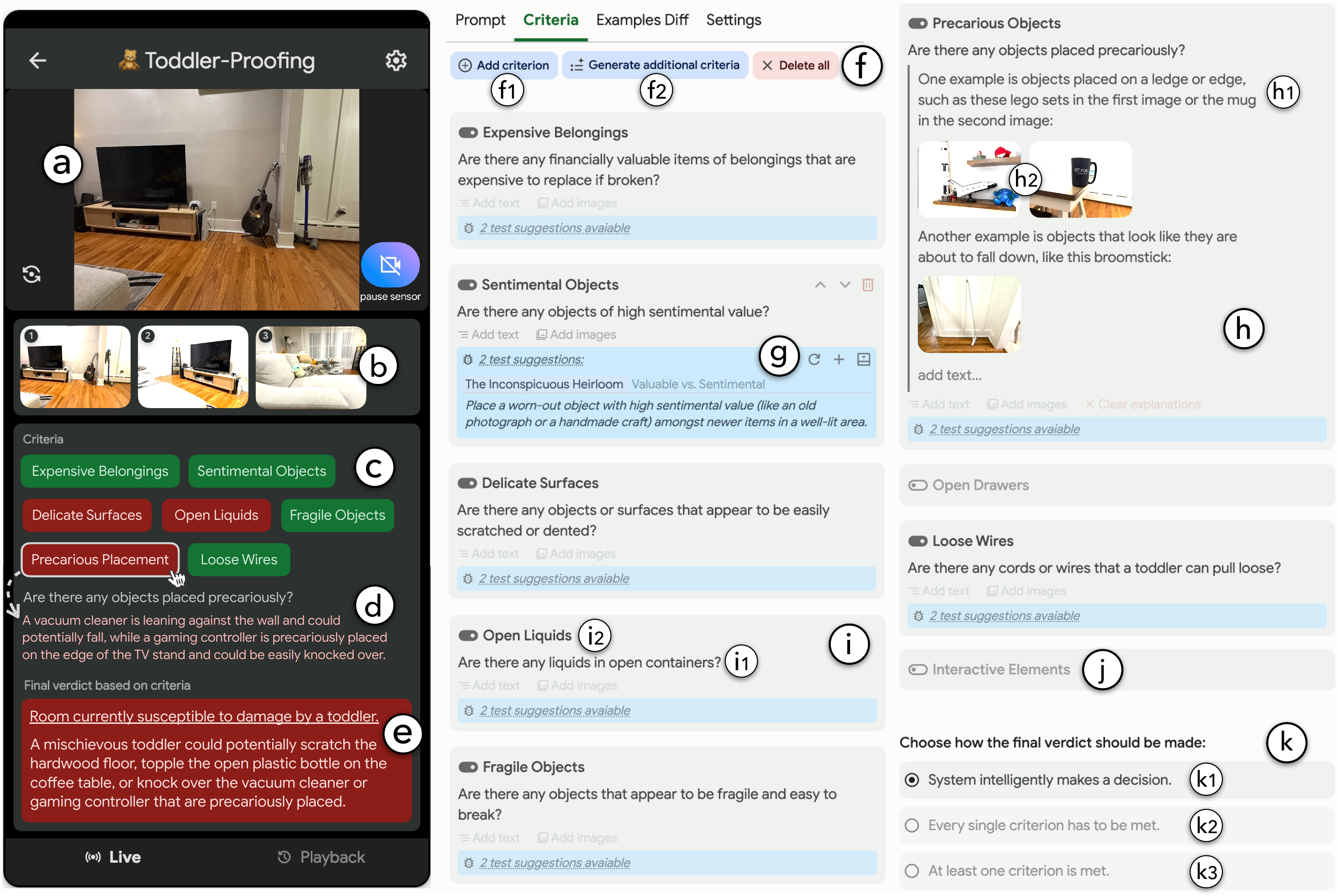}
\vspace{-7mm}
\caption{
\systemname' main user interface for formulating and curating criteria that govern a sensor's behavior. For a high-level sensing task (in this case ``tell me if my toddler might damage something''), the system operates as follows: At each time interval, the sensor evaluates all criteria (c) simultaneously using frames captured by a live camera (a) over the past few seconds (b). The result for each criterion is displayed as a green or red chip, indicating a positive or negative outcome, respectively. Users can click on a chip to view detailed results for that specific criterion, which is the description from the MLLM's interpretation of the scene (d). The sensor then synthesizes these individual results to make an informed final decision regarding the original sensing task (e).
In the criteria editor (f), users have the option to either add their own criterion (f1) or have \systemname automatically generate criteria (f2) based on the initial sensing task and the live camera view. To modify an existing criterion (i), users can update its description (i1), and \systemname will automatically generate a title (i2) for display in the Live sensor view (c).
Furthermore, users can 1) add additional text (h1) or visual examples (h2) to explain a criterion (h) based on their personal context; 2) review \systemname-generated suggestions for testing a specific criterion (g); 3) temporarily enable or disable a criterion (j); and 4) configure how the sensor reaches its final verdict (k), ranging from allowing the MLLM to make an intelligent decision based on all criteria results (k1), to rule-based combinations using boolean logic (k2, k3).\looseness=-1
}
\label{fig:teaser}
\vspace{-2mm}
\end{figure*}


The proliferation of smart home technologies has popularized domestic sensing and monitoring \cite{woo_user_2015,hwang_smart_2012,jakobi_evolving_2018,yun_potential_2023,laput_ubicoustics_2018}. \revision{Among these,} smart camera sensors, offered by \revision{major} consumer electronics companies, \revision{enable} individuals to observe and react to events \revision{or activities} within and around their homes, \revision{thereby} enhancing safety and awareness. These devices leverage advances in computer vision (CV) and machine learning (ML) to identify and classify objects, distinguish between human and animal activities, or \revision{recognize} specific events like package deliveries \cite{ring_security_2024}.
However, the current generation of smart camera sensors have limitations in their adaptability and configurability \cite{nam_indoor_2016,pierce_sensor_2020,ye_progesar_2022}. While capable of detecting predefined events, these systems do not readily support nuanced needs and priorities of individual users. This can lead to an \revision{excess} of irrelevant notifications, missed critical events, and a mismatch between sensor capabilities, environmental conditions, and user needs.\looseness=-1

In contrast to classical \revision{CV and ML models}, recent multimodal large language models (MLLMs) integrate vast amounts of world knowledge and can understand and reason across a mixture of visual and textual inputs, much like humans \cite{fei_towards_2022,li_multimodal_2024}. This advancement \revision{unlocks} new \revision{opportunities} for end-users to \textit{define} personalized, MLLM-driven sensors capable of reasoning about complex situations \revision{(referred to hereafter as ``AI sensors'' for brevity)}--a user could \revision{describe} a desired sensing task in natural language, with MLLMs analyzing camera image frames and providing responses within seconds.\looseness=-1

To illustrate, today's standard CV/ML classifier-based sensors are typically trained on domain-specific data \cite{kastrinaki_survey_2003} and limited to \textit{lower-level} specific sensing objectives (e.g. \revision{``are there unusual movements near the window'',} ``is there a pet in the kitchen''). In contrast, MLLM-driven AI sensors offer significantly greater flexibility with their ability to \textit{reason} about \textit{higher-level} problems (e.g. ``alert me if my toddler is getting into trouble in the living room'', ``is there something out of place in the backyard?''). These reasoning capabilities may enable sensors to interpret the world in ways that would otherwise be challenging for traditional CV/ML classifiers. \revision{However}, the seemingly infinite scope of tasks that can be accomplished with MLLMs also means that the semantic space of sensors users could define is also much broader than before, as they can now define tasks that are potentially more abstract, high-level, and \textit{subjective} (e.g. ``is my room messy?'').

In a formative study, we confirmed users' desire for AI sensor personalization and configurability, but also discovered that the system must better support users in defining and communicating personal \textit{criteria} and \textit{constraints} (e.g. ``is there something out of place in my room'' [ignore the messy power cords]). 
This need was particularly \revision{pronounced} for higher-level sensors with multiple possible semantic interpretations. For instance, a toddler ``getting into mischief'' might mean unrolling an entire roll of toilet paper in one household versus drawing on their sibling's face in another. 
Additionally, users often struggled to generalize beyond their immediate, personal settings (e.g., their own room) and \revision{found it challenging to anticipate future scenarios and variants of those settings where their sensors might behave unexpectedly.}
\revision{These findings suggest that} effective AI sensor definition requires systems to anchor \textit{more} \revision{closely to users'} own criteria, while \revision{simultaneously} making them aware of \textit{other}, unanticipated scenarios.

To address these AI sensor definition opportunities and challenges, we developed \systemname, a system that empowers users to create and test real-time sensors powered by MLLMs. 
Beyond allowing users to create ad-hoc sensors via natural language prompting, \systemname leverages the reasoning capabilities and world knowledge of MLLMs. Specifically, users can (1) ask \systemname to break down the sensing problem into \textbf{automatically-generated relevant criteria}, (2) \textbf{manually define their own criteria}, and (3) \textbf{test and debug} multiple criteria \revision{simultaneously} in real time. Furthermore, users can (4) ask \systemname to \textbf{generate new criteria not yet considered} based on frames from their video stream as positive and negative examples and also (5) \textbf{``future-proof''} their sensors by asking \systemname to suggest future situations that may lead to model failures, along with actionable tests for the user to try.
Finally, users can \revision{configure} how the final verdict is determined when considering the collective criteria, with options for an LLM-generated decision or rule-based combinations using boolean logic.\looseness=-1

In a user study with 12 participants, we compared \systemname to a baseline condition where users iterated on a single prompt without structured assistance. We found that \systemname significantly increased users' sense of control, understanding, and communication with the model. Specifically, \systemname enabled participants to decompose the AI sensor definition problem into lower-level criteria and explore them in parallel, granting them greater control and systematic insights. Additionally, \systemname' reasoning capabilities also helped offset users' own limitations and ``blind spots,'' by making them aware of potential failure modes and edge cases, and \revision{surfacing context-specific criteria they hadn't considered}.
Finally, our study also highlighted how certain idiosyncrasies of MLLMs, such as hallucinations or ``flickering'' textual responses, impact sensor performance and \revision{perceived} reliability.

Notably, participants used the \systemname tools for purposes \textit{beyond} addressing model limitations: by enabling users to decompose sensors into bite-sized criteria and test them, \systemname helped users to more easily elicit their own personal requirements, debug the sensor through isolating sub-components of logic, and better understand the model's capabilities and limitations. These needs will likely persist even as models continue to improve in the future. Collectively, these findings contribute to the design of future user-defined sensing systems supported by MLLM-powered reasoning.

\vspace{3mm}
\noindent\textbf{Contributions. }This paper makes the following contributions:

\begin{itemize}[leftmargin=0.12in,topsep=5pt]
\item \textbf{Formative study} (n=6) confirming the potential and desire for end-user AI sensor definition, alongside challenges with open-ended prompting.

\item \textbf{Design goals for AI sensor definition}: support for control over sensors' behavior, criteria elicitation, expression of personal constraints, and systematic testing and debugging. 

\item \textbf{The {\systemname system}} for more effective AI sensor definition through automatic and manual generation of criteria, debugging of individual criteria, example-driven specification, and actionable test cases.

\item \textbf{Formal user study} (n=12) finding that \systemname significantly increased users' sense of control, understanding, and communication with the model, through enabling users to focus more on formulating and debugging sensor requirements rather than on prompt phrasing. Beyond this, \systemname' reasoning capabilities also helped users consider failure modes beyond their own immediate situations, and helped users address their own ``blind spots'' by exposing criteria they didn't already think of. Finally, we discovered how MLLM idiosyncrasies (``flickering'' and hallucinations) affected the sensor creation process.

\end{itemize}

\section{Related Work}\label{sec:rw}

\subsection{Intelligent and DIY Visual Sensing}
The concept of general-purpose, do-it-yourself (DIY) sensing has long been considered the ultimate goal of ubiquitous computing, particularly in smart home environments \cite{woo_user_2015,hwang_smart_2012,jakobi_evolving_2018,yun_potential_2023,laput_ubicoustics_2018}. Many have aspired for sensing systems that end-users can intuitively customize \cite{noura_natural_2020,laput_synthetic_2017}, yet current commercial smart home sensors still fall short of this vision. While affordable and accessible, they are typically highly specialized \cite{scott_preheat_2011,klingensmith_hot_2014,kuznetsov_upstream_2010}, and require users to invest significant time and effort in learning and creating custom workflows based on their outputs \cite{shi_knock_2018}. Moreover, these sensors usually produce low-level data that cannot directly answer users' high-level questions \cite{ding_sensor_2011,hong_evidential_2009}. For example, a door sensor might indicate the door's open/close status but cannot explicitly inform users whether their children have left or arrived home \cite{laput_zensors_2015}.\looseness=-1

Much prior research has been directed at closing this gap between \textit{what can technically be sensed} and \textit{what users are actually interested in knowing} \cite{ding_sensor_2011,li_contextual_2024}, particularly in the visual sensing domain. Earlier work focused on leveraging ``human intelligence'' through online marketplaces such as Amazon Mechanical Turk \cite{amazon_amazon_2024}. For example, VizWiz \cite{bigham_vizwiz_2010} and VizLens \cite{guo_vizlens_2016} had crowd workers answer visual questions of photos taken from smartphones. 
Follow-up efforts, like VATIC \cite{vondrick_efficiently_2013} and Flock \cite{cheng_flock_2015}, utilize crowd-labeled data to subsequently train ML models.
These efforts culminated in the Zensors system \cite{laput_zensors_2015,guo_crowd-ai_2018}: initially, Zensors relies on human intelligence to directly answer users' sensing questions, such as ``Is there parking spots available?'' or ``How orderly is the line?'' Over time, it uses this human-labeled data to train CV models, ultimately automating the sensing task by replacing human input with model predictions.\looseness=-1

Though providing a general-purpose sensing solution, Zensors' effectiveness and applicability remains constrained by the capabilities and limitations of traditional CV models. Additionally, users may struggle to debug or customize sensors \cite{laput_zensors_2015}, especially when outputs differ from expectations. In this work, we explore the potential of replacing crowd-ML hybrid sensing backbone with MLLMs, offering two distinct advantages: 1) MLLMs, with their reasoning capabilities, can directly explain their \textit{thought processes}, aiding users in sensor debugging and understanding system capabilities and limitations, unlike previous methods where the rationale used by crowd workers is obscured by the later CV model; 2) MLLMs handle multimodal inputs, allowing users to define sensors using not only natural language but also direct visual examples, offering greater flexibility to express their unique personal contexts and needs.


\subsection{LLM Prompting and Requirement Articulation}
In the emerging paradigm of end-user-defined AI sensors, users are now responsible for ensuring their sensors operate as intended, largely through crafting effective prompts. 
Indeed, prompting has become crucial for crafting effective input instructions to guide Large Language Models (LLMs) in generating desired outputs \cite{liu_we_2024,zamfirescu-pereira_why_2023,zamfirescu-pereira_herding_2023,liu_pre-train_2023,bach_promptsource_2022,qian_evolution_2024}, and has dramatically democratized and accelerated AI prototyping across various use cases and domains \cite{jiang_promptmaker_2022,jiang_genline_2021,petridis_anglekindling_2023,petridis_promptinfuser_2023,petridis_constitutionmaker_2024,petridis_situ_2024,liu_tool_2023,russell_sensemaking_2024,liu_what_2023,kahng_llm_2024,kahng_llm_2024-1,liu_selenite_2024,lee_benefits_2023}. However, prompting remains a challenging and ambiguous task, particularly for users without technical expertise in LLMs. Common challenges include struggling with finding the right phrasing for a prompt, selecting appropriate demonstrative examples, experimenting with various hyper-parameters, and evaluating the effectiveness of their prompts \cite{jiang_promptmaker_2022,denny_prompt_2024,zamfirescu-pereira_why_2023}. Consequently, they may waste time on unproductive strategies, such as making trivial wording changes \cite{parnin_building_2023,nguyen_how_2024}. This issue is further exacerbated when the task becomes more complex, involving multiple facets and requirements that need to be addressed within a single prompt \cite{jiang_followbench_2024,zhou_instruction-following_2023}. Similarly, we observed in the formative study that participants often haphazardly make ad hoc revisions to their sensing prompts in response to previous outputs, without a clear understanding of what needs improvement.

Akin to the concept of \textit{requirement engineering} in software engineering \cite{ko_state_2011}, where humans define the desired outcomes and behavior of a program, often including expected inputs and outputs \cite{lamsweerde_requirements_2009}, recent research has shown that explicitly and clearly stating \textit{requirements} within prompts is an effective strategy for systematically improving them \cite{desmond_exploring_2024,mesko_prompt_2023,schmidt_towards_2024}. However, articulating clear and complete requirements is a known challenge \cite{liu_what_2023,nam_using_2024,liu_reuse_2021,nguyen_how_2024,liu_unakite:_2019,hsieh_exploratory_2018}, even for experts who need multiple iterations to refine them \cite{shankar_spade_2024}. Poorly defined requirements frequently lead to program failures \cite{millett_software_2007,denny_prompt_2024}. While tools like EvalLM \cite{kim_evallm_2024}, SPADE \cite{shankar_spade_2024}, and EvalGen \cite{shankar_who_2024} have explored extracting user requirements from prompts to support prompt evaluation, there is limited emphasis on assisting users in effectively communicating requirements during prompt construction.
In this work, we address this gap by prioritizing requirements as first-class entities in the form of \textit{criteria} that govern the sensor behaviors.
Rather than having end-users juggle the intricacies of prompt writing, we abstract these mechanics of raw prompt away from them, and instead direct their attention on defining criteria, thereby simplifying the task. Additionally, we offer user support through features such as auto-generating criteria based on common sense or user-provided visual examples, further assisting users in translating their personal contexts and preferences into criteria.\looseness=-1


\subsection{Interactive Model Refinement Through User Feedback}

Prior work has explored various interactive systems that allow users to provide feedback to refine future model outputs. Programming-by-example tools enable users to provide input-output examples, with the system generating a function that fits these examples \cite{chen_multi-modal_2020,verbruggen_semantic_2021,zhang_interactive_2020}. Similarly, recommender systems allow users to steer outputs through limited feedback \cite{petridis_tastepaths_2022,bostandjiev_tasteweights_2012,liu_crystalline_2022}, such as adjusting a 2D plane to influence movie recommendations \cite{kunkel_3d_2017}. Teachable Machines also offer an interactive approach, allowing users to train ML models by supplying labeled examples, with real-time feedback facilitating iterative refinement \cite{carney_teachable_2020}. More recently, systems and methods like ConstitutionMaker \cite{petridis_constitutionmaker_2024} and ConstitutionalExperts \cite{petridis_constitutionalexperts_2024} leverage LLMs to translate natural language feedback and critique into high-level \textit{principles} (similar to the sensor criteria in our work), enabling conversational, human-like interactions to steer the model.
However, previous approaches often integrate natural language principles directly into prompts to steer model outputs \cite{petridis_constitutionmaker_2024,liu_what_2023}, making it difficult for users to understand how individual principles perform or how editing one principle might inadvertently affect the efficacy of others. In contrast, in \systemname, each criterion targets a single, specific aspect of the overall problem. Users adjust one criterion at a time and receive results specific to that criterion without having to disentangle insights from a mix of different factors bundled into a larger prompt response, and without fearing that adjustments might affect other criteria.

\section{Formative Study \revision{\& Design Goals}}\label{sec:formative-study}

To understand the opportunities and potential challenges for user-specified AI sensors, we conducted a formative study with six professional designers (age range: 29-42, 3 female and 3 male) from a large technology company, where they were asked to brainstorm AI sensor use cases and create their own sensors with a prompt-based prototype. \revision{Based on the findings from this study and insights from prior research, we identified a set of design goals for \systemname.}

\subsection{Setup}

\subsubsection{Procedure} 
The overall outline of the formative study was as follows:
(1) Participants spent 5 minutes individually brainstorming potential use cases for visual-based personal sensors, documenting their ideas and initial thoughts.
(2) Participants were then shown the prompt-based prototype they would be using to build their sensors (see Figure \ref{fig:baseline-prompt-editor}).
(3) Participants spent 40 minutes individually creating one to two sensors of their choice. While they prototyped, they were asked to take notes on how they iterated over their prompt and take screenshots.
To ensure ecological validity, participants were asked to create sensors in their homes.
(4) For the last 10 minutes, the facilitators led a group discussion with the participants to learn about their experience building sensors.

\subsubsection{Prototype} The prototype (Figure \ref{fig:baseline-prompt-editor}) provided a basic interface for creating MLLM-powered sensors.
The prototype allowed them to input prompts (Figure \ref{fig:baseline-prompt-editor}-d) and set an interval for execution using their laptop camera.
The prototype displayed the MLLM's output for each execution (Figure \ref{fig:baseline-prompt-editor}-e), as well as the corresponding input image (Figure \ref{fig:baseline-prompt-editor}-c).
Finally, participants could view the history of all the sensor inputs and outputs.\looseness=-1

\subsection{Findings}

\subsubsection{The opportunity for personal AI-powered sensors}
All participants were quite excited about the possibility of building their own personal sensors.
P3 explained, \userquote{Traditionally, I had to buy a product built for a task...being able to set up my own sensor - that's the new part.}
He then went on to describe that normally the company selling the sensor determines how the sensor behaves, but now he is able to create and customize the sensor to his own requirements.\looseness=-1

Participants brainstormed a total of 47 personal sensors they would use in their daily lives.
They ideated \textit{reminder} sensors which would detect the last time they watered the plants, exercised, or took their vitamins.
There were also \textit{aggregative} sensors that detected: ``how much family time are we spending,'' ``how much time did I spend practicing Korean,'' ``what food have I eaten during the day to manage health conditions.''
In addition to these more reflective sensors, participants also had many ideas for more \textit{urgent} sensors, including: ``let me know if my landlord is at the door,'' ``tell me if my pork chop is getting burnt,'' and ``tell me if there's a leak at my water heater.''
Finally, participants also had ideas for sensors that detected \textit{new events}, such as ``how many new people did I meet last week'' and ``tell me when I see something I have never seen before.''\looseness=-1

\begin{figure}[t]
\centering
    \includegraphics[width=0.78\linewidth]{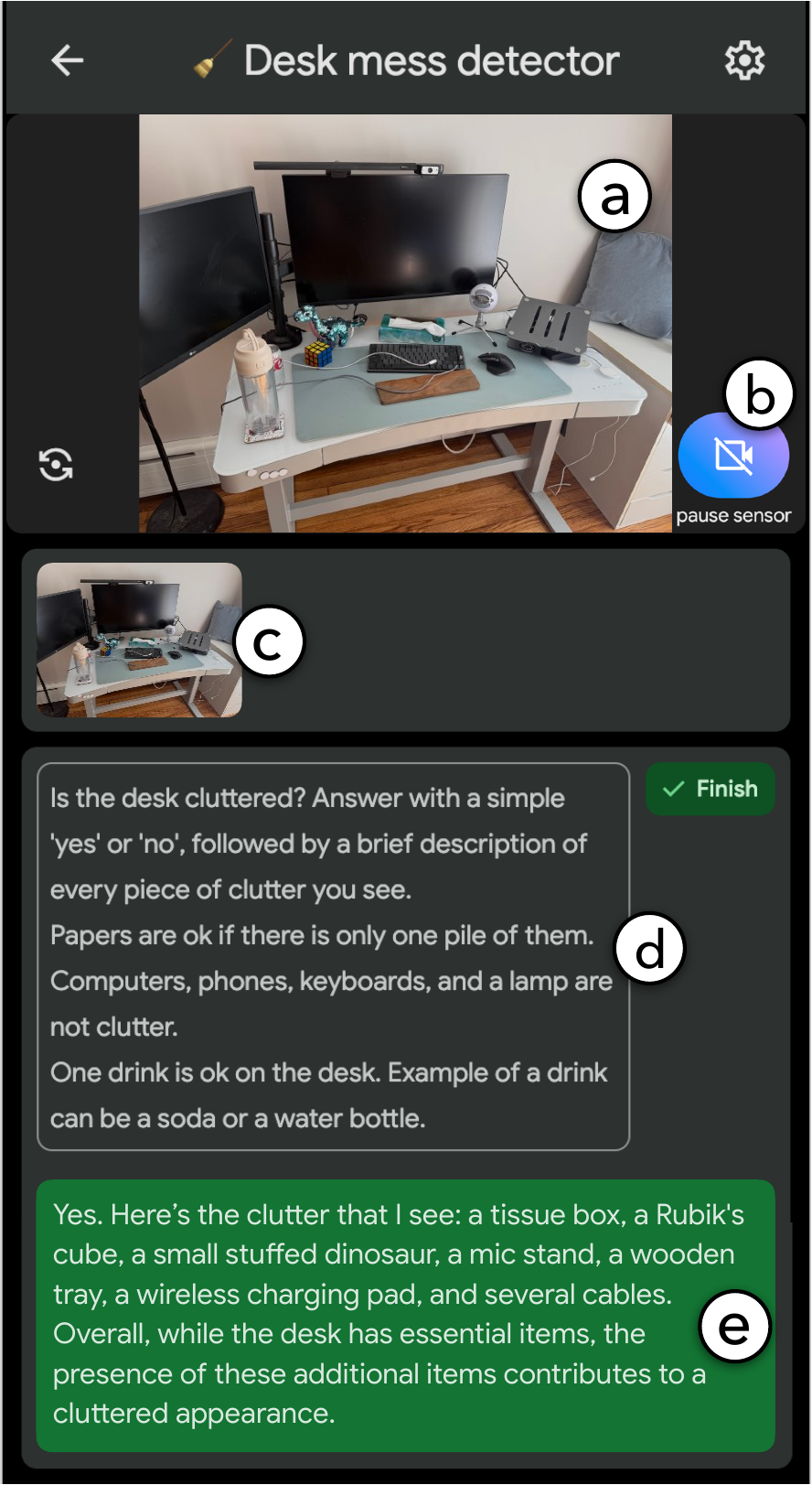}
  \vspace{-3mm}
  \caption{\textbf{The prompt-editor authoring tool} participants used in the formative study, as well as the baseline condition in the user study (see Section \ref{user-study-procedure}). Users can view what their camera currently sees (a) and pause and play the sensor (b). They can edit their prompt (d) and then view its latest result (e) as well as the corresponding input image it was run on (c).}
  \label{fig:baseline-prompt-editor}
  \vspace{-5mm}
\end{figure}

\subsubsection{Challenges in creating prompt-powered sensors}
A common strategy participants employed to steer their sensors was to write criteria to specify when the sensor should output a label.
For example, P6 built a sensor to determine when his plant needs fertilizer. Their prompt was initially: ``\promptText{Does this plant need fertilizer? Explain why}.''
The outputs from this prompt were a bit vague and seemed unjustified, like: ``\promptText{No. The plants look healthy}.''
So, they then appended the criterion: ``\promptText{Look for signs of yellowing and discoloration}'' which steered the model toward producing more relevant and informative explanations.
Similarly, P5 was building a sensor that detects if any chores needed to be done in her living room, and to further steer the sensor, she appended a criterion to check for any trash in the bins.
By adding these criteria, participants were able to guide the model toward better decisions and explanations from their sensors.\looseness=-1

\revision{However, steering the model using criteria presented several challenges.}
First, as participants added more criteria and clauses to their prompts, they became unwieldy; it became \revision{evident that the model neither adhered to nor explicitly checked all the specified criteria.}
Additionally, these convoluted, multi-clause prompts \revision{left participants uncertain about how to modify their criteria or adjust the prompt as a whole.}
For this reason, our first design goal (\textbf{D1}) was to support more \textbf{precise control over a sensor's behavior} and enable users to incorporate and test their criteria individually.

\revision{Identifying useful criteria was another significant hurdle for participants.}
\revision{For instance,} P2, who built a sensor to detect if he was eating unhealthy snacks, \revision{struggled to} define \revision{a comprehensive} set of qualities that \revision{characterized} the snacks they personally \revision{considered} unhealthy.
Therefore, our second design goal (\textbf{D2}) was to \textbf{accelerate requirement elicitation} by \revision{offering users} an initial set of common-sense criteria to peruse \revision{and refine}.\looseness=-1

Next, \revision{participants sometimes struggled to articulate} particular criteria in their prompts.
For example, P3 made a sensor to detect if his living room was messy. He had a few decorative items on the couch, including a few unique pillows and a throw blanket \revision{featuring} a pizza design.
\revision{In his prompt, P3 specified the criterion:} ``\promptText{look for a remote or wrappers on the couch, not the pillows or blanket}.''
However, despite adjusting the wording of this criteria, the sensor continued to classify the couch with the throw blanket as messy; P3 \revision{expressed a desire to illustrate this particular criterion with an image, emphasizing that the distinctive throw blanket should not be considered part of the mess.}
\revision{This led to} our third design goal (\textbf{D3}): to provide \textbf{flexible ways for users to communicate their personal context}, \revision{whether} by enabling them to \revision{use images to clarify criteria} or by helping them verbalize more nuanced conditions.

Finally, two participants expressed doubts in their sensors' future performance.
P3 \revision{described experiencing a sense of} ``tunnel vision'' with his tests, as his experiments were limited to removing and adding nearby objects to the couch and coffee table in his living room.
It was hard for P3 to \revision{step back from his immediate} physical context and \revision{envision realistic changes his living room might undergo over time}.
Thus, our final design goal (\textbf{D4}) was to \textbf{scaffold testing} and support users in future-proofing their sensors.\looseness=-1

\subsection{Summary of Design Goals}
In summary, we postulate that an effective system that helps end-users create flexible and intelligent AI sensors should support:

\begin{itemize}[leftmargin=0.12in,topsep=5pt]
\item \textbf{D1}: \textbf{Enabling precise control over the sensor behavior via fine-grained criteria}. Users should be able to author and adjust criteria individually and assess the sensor's performance for each one, without needing to revise an entire prompt.

\item \textbf{D2}: \textbf{Accelerating requirement elicitation by bootstrapping common-sense criteria}. The system should assist users in getting started by generating relevant, common-sense criteria automatically.

\item \textbf{D3}: \textbf{Providing flexible ways to communicate personal context}. Users should be able to express their specific context via text, as well as visually. The system should also help users identify their more nuanced criteria.

\item \textbf{D4}: \textbf{Scaffolding testing and debugging of criteria}. The system should offer tools that allow users to test, isolate, and debug each criterion separately, enabling users to address future scenarios that might confound their sensor, incrementally improve the sensor's accuracy, and understand how each criterion impacts overall performance.
\end{itemize}

\section{The \systemname System}

We begin by presenting a usage scenario that demonstrates the core functionalities of \systemname. This example incorporates key use cases identified in our formative study, along with specific criteria curated by participants in the subsequent user study.

\subsection{Example Usage Scenario}
\exampleScenarioUser was concerned about her toddler ``getting into trouble'' in her home, as she has often observed him playing with her purses or breaking her valuable belongings. She couldn't find any commercial sensors specifically designed to monitor this sort of situation, so she decided to create a custom one using the \systemname platform. To get started, \exampleScenarioUser entered her request as ``tell me if toddler might damage something'' into the system. \systemname automatically generated a basic LLM prompt based on that request: ``\promptText{Is the toddler likely to damage something valuable in this room? Answer with `Yes' or `No', and provide a brief explanation}.'' \exampleScenarioUser then positioned her webcam to provide a clear view of her living room space where her toddler frequently played. Within seconds, the custom ``Toddler Check'' sensor was operational: it is configured to run every three seconds (a \revision{default} frequency easy for users to test and debug the sensor prompt) using the latest three frames (\fig\ref{fig:teaser}-b) from the camera feed (\fig\ref{fig:teaser}-a, which takes one frame per second) and the prompt mentioned previously, providing continuous assessment of the room.\looseness=-1

However, \exampleScenarioUser quickly realized that the sensor considered her stack of clothes damageable, even though she is fine and well-accustomed with her toddler playing with her laundry. The sensor also did not notice her sacred wedding photo that her toddler often attempts to reach. Despite her efforts to tweak and add more clauses, it was unclear if these adjustments were consistently improving the responses. As a result, \exampleScenarioUser clicked the ``Generate criteria'' button (\fig\ref{fig:teaser}-f2), and \systemname automatically produced several criteria (e.g., \fig\ref{fig:teaser}-i) based on the initial request and the environment (e.g. ``\promptText{Delicate Surfaces},'' ``\promptText{Sentimental Objects},'' ``\promptText{Fragile Objects}''). To try out these criteria, she could view the real-time results for each criterion as small green or red chips in the sensor's Live View (\fig\ref{fig:teaser}-c), clearly indicating whether the room passed or failed at each criterion check. \exampleScenarioUser could click on any chip to read a brief explanation, helping her understand what the model ``saw'' in the image (\fig\ref{fig:teaser}-d). She realized that the sensor was consistently missing  ``\promptText{Precarious Objects},'' so she modified the criterion to include a few image examples, such as objects placed on a ledge or edge or those that could potentially fall (\fig\ref{fig:teaser}-h). She also added her own ``\promptText{Open Liquids}'' criteria when she saw the open water bottle on the coffee table.\looseness=-1

\begin{figure*}[t]
\vspace{-2mm}
\centering
\includegraphics[width=0.7\textwidth]{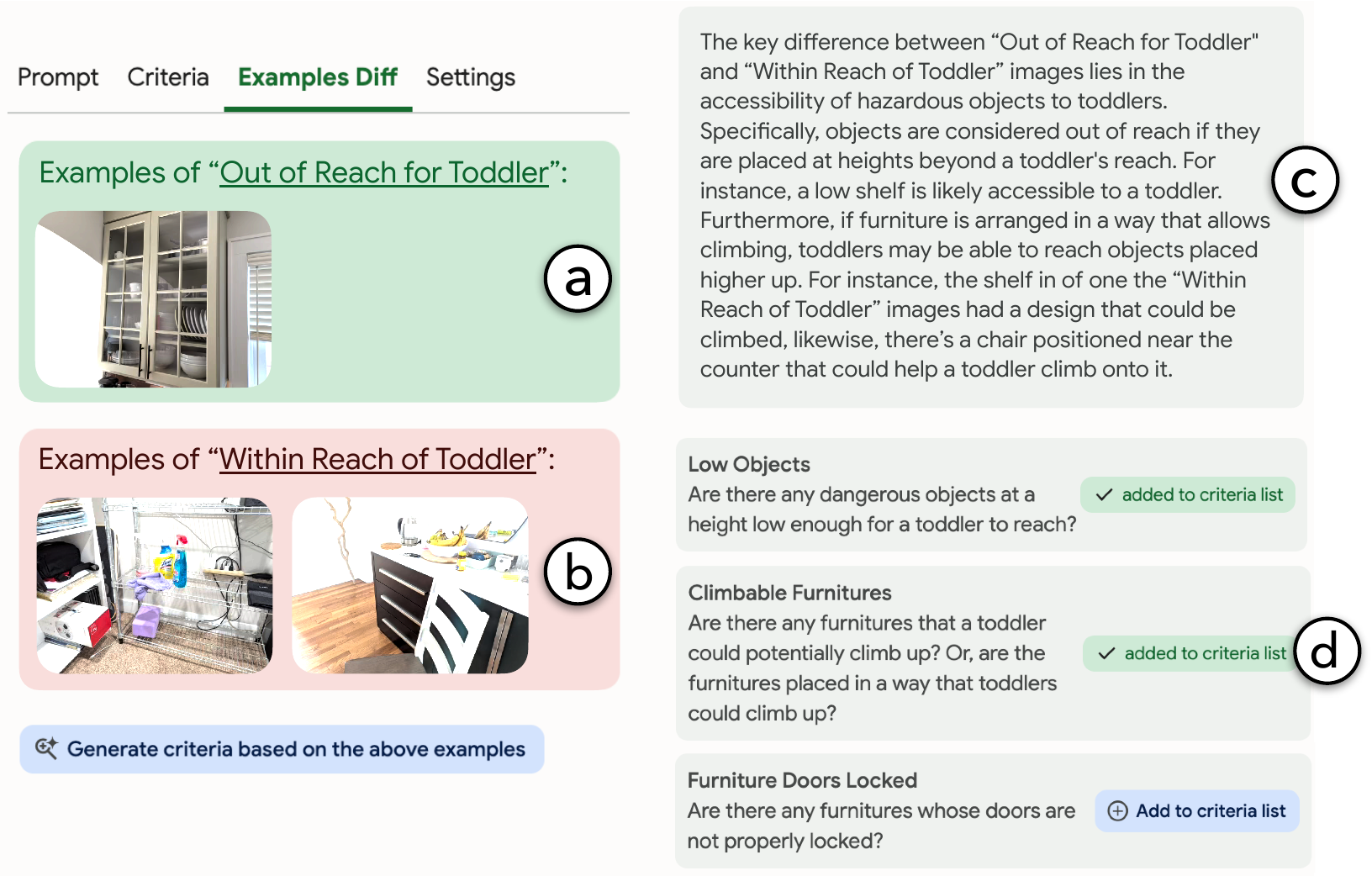}
\vspace{-3mm}
\caption{
The Examples-diff feature in \systemname allows users to provide visual examples for each answer category (a and b) related to the high-level sensing task. \systemname then \revision{presents} its reasoning process (c) and any additional criteria generated from that reasoning (d). Users can choose whether to incorporate these criteria into the main list.
}
\vspace{-2mm}
\label{fig:examples-diff}
\end{figure*}

Later, \exampleScenarioUser \revision{found it challenging to articulate} the concept of ``books the toddler shouldn't tear up'' in words, because it was hard to verbally distinguish between her own books from the toddler's children's books, so she tried the Example-Diff feature (See \fig\ref{fig:examples-diff}). She pointed the webcam to capture these items, then from the Playback view (that contains the history of all sensor runs), she selected image frames from that earlier webcam footage: a set of her books, and a set of children's books. To \exampleScenarioUser's surprise, the system then suggested additional criteria she hadn't considered. \revision{For instance, it noted that adult books were less likely to have large pictures on the covers and often featured paper dust jackets. \exampleScenarioUser incorporated these into her existing set of criteria, recognizing that the system had highlighted aspects she might have otherwise overlooked.}\looseness=-1

\subsection{System and User Interface Design}

We now discuss how the various \systemname features are designed and implemented to support the design goals. \systemname serves as a platform for end-users to create and experiment AI sensors powered by multimodal foundation models. These sensors are therefore hardware-agnostic, and can eventually be deployed on any existing hardware with a camera feed. It is important to note that challenges related to actual sensor deployment and maintenance fall outside the scope of this work.

\subsubsection{D1: Enable precise control over the sensor behavior via fine-grained criteria.}

In \systemname, a criterion functions as an atomic unit of reasoning, and is essentially a minimal prompt that targets a single, specific aspect (e.g., ``\promptText{Are there any open outlets that are not properly covered?}'') of the overall sensing problem (e.g., ``Is this room safe for a toddler?''). This level of granularity \revision{ helps prevent} MLLMs from overlooking details from the input image frames, a common \revision{challenge} identified in our formative study and \revision{corroborated by} previous research on complex or multi-faceted prompts \cite{liu_we_2024}. In addition, by isolating individual criteria, users can more easily interpret results \revision{(\fig\ref{fig:teaser}-d)} without having to disentangle insights from a mix of different factors bundled into one larger prompt response, which can be cognitively demanding.

Creating a new criterion is as straightforward as writing \revision{a question in natural language}, lowering the barrier for users without technical expertise to start making sensors. Users can create a criterion by clicking the ``Add Criterion'' button at the top of the Criteria tab (\fig\ref{fig:teaser}-f1). Once they type in their question, the system automatically generates a concise name for the criterion, making it easy to identify and parse among other criteria when viewing results in the Live view (\fig\ref{fig:teaser}-c).\looseness=-1

Once a criterion is created, \systemname~ \revision{evaluates} it against the scene during subsequent runs. Specifically, we instruct the MLLM through system instructions to always generate a brief description regarding the criterion, \revision{accompanied by} a valence \revision{that reflects} the semantic outcome of the evaluation. This valence, indicating whether the scene has passed the check or if issues have been detected, is visually represented in the Live view using color-coded chips: green signifies a positive outcome \textit{with no issues}, while red indicates a negative one \textit{with potential issues that require user attention}.\footnote{We acknowledge the accessibility concerns for color-blind users with this color scheme and plan to address them in future system iterations.} 
For instance, regarding the question 
``\promptText{Are there any objects placed precariously?}'' \revision{(\fig\ref{fig:teaser}-d)}, if the model indeed identifies objects that are placed precariously (such as those mentioned in the Example Usage Scenario, see \fig\ref{fig:teaser}-h), the valence will be negative (hence red criterion chip). \revision{Conversely,} if the model considers \revision{all} objects to be properly placed, the valence will be positive (hence green criterion chip).
\revision{These} color-coded chips enable users to quickly assess a sensor's behavior at a high-level, and detect potential issues at a glance (\fig\ref{fig:teaser}-c). \revision{If} a user notices something out of the ordinary—such as a red chip indicating an unmet condition—they can click on the chip to access a more detailed \revision{explanation of the issue} (\fig\ref{fig:teaser}-d).\looseness=-1

Furthermore, each criterion operates independently from others, ensuring that changes to one do not inadvertently affect the performance of others, unlike the behavior observed when iterating on a single, large prompt in the formative study (\fig\ref{fig:baseline-prompt-editor}). Users also have the option to deactivate a criterion if they \revision{wish to suppress} its results temporarily. \revision{Once satisfied with the performance of a particular criterion}, they can \revision{proceed} to the next, resulting in a more systematic and tractable debugging process.

At each time interval, all active criteria are executed in parallel, each yielding its own result. These individual results are then aggregated and fed into a subsequent prompt, which is tasked with reasoning through each criterion result and then formulating an informed final verdict \revision{accompanied by} an explanation. Our informal testing suggests this divergent-then-convergent approach effectively ensures the sensor considers all the aspects that the user specified, while intelligently analyzing and prioritizing these aspects based on common sense (\fig\ref{fig:teaser}-k1). 
Alternatively, users have the flexibility to instead use Boolean logic for the final verdict. 
\revision{For instance, they can configure the system to require all criteria to be met for a positive outcome (AND) (\fig\ref{fig:teaser}-k2) or allow a positive verdict if at least one criterion is satisfied (OR) (\fig\ref{fig:teaser}-k3).}

\subsubsection{D2: Accelerating requirement elicitation by bootstrapping common sense criteria.}

To further reduce user friction, \systemname can also automatically generate relevant criteria (\fig\ref{fig:teaser}-f2) by leveraging the extensive world knowledge and reasoning capabilities of today's foundation models. Specifically, we prompted the model to generate criteria that are contextually grounded in the user's sensing task and environment, mirroring what a human would typically use to evaluate the task. These auto-generated criteria \revision{serve} a useful starting point, \revision{particularly} for users \revision{less} unfamiliar with the intricacies of \revision{their tasks}. Unlike previous systems that aimed \revision{for comprehensive coverage upfront} (e.g., Selenite \cite{liu_selenite_2024}), we intentionally limit the number of criteria generated \revision{per turn} to four. This strategy \revision{avoids} overwhelming users with too many suggestions at once, thereby maintaining their cognitive bandwidth and agency to define criteria based on their unique personal context. \revision{Users still can, however, generate additional sets of criteria, each guaranteed to differ from the existing ones,} empowering them to explore as many \revision{perspectives} as they wish without being locked into a single set of \revision{recommendations}.\looseness=-1

\subsubsection{D3: Provide flexible ways to communicate personal context.}

In our formative study, participants often found it challenging to define criteria precisely using only natural language. Often, a more intuitive approach for them was to provide concrete visual examples, such as annotated images, \revision{in combination with} textual descriptions, hoping that the system would infer the intended meaning more effectively. 
To address this challenge, \systemname enables users to communicate their criteria through not only text but also images and annotations that represent their expectations when needed (\fig\ref{fig:teaser}-h). This flexibility empowers users to select the modality they naturally gravitate towards depending on the circumstances, and can be particularly useful when describing visual or abstract features.  Behind the scenes, \systemname combines all content under a single criterion together to steer the MLLM's understanding and subsequent responses.\looseness=-1

Furthermore, it could often be the case where users have an abstract idea of what they want a sensor to monitor but struggle to articulate these ideas into specific criteria. \revision{To address this}, \systemname~ \revision{offers the} Examples-Diff feature, which \revision{transforms} labeled images into clear, actionable criteria (\fig\ref{fig:examples-diff}). For each sensing task, \systemname automatically generates distinct categories of possible answers based on the original sensing task, \revision{while also allowing users to} customize these categories to their liking. Users can then select representative image frames from the sensor's history to illustrate each category \revision{(\fig\ref{fig:examples-diff}-a\&b)}. 
Behind the scenes, we leverage the visual reasoning capabilities of MLLMs, and instruct the model to first ``reason through the provided images and think carefully about their differences as well as the subtleties the user is trying to convey through these examples'' and then generate ``actionable criteria that are absent from or inadequately represented by existing criteria.'' We present the model's reasoning process \revision{(\fig\ref{fig:examples-diff}-c)} and the newly generated criteria \revision{(\fig\ref{fig:examples-diff}-d)} to the user, who can \revision{review and decide} whether to incorporate them into the main criteria list.

This feature is especially useful in boundary cases where it is challenging for users to manually differentiate and develop effective criteria for the model to understand and process. Additionally, when users struggle with inspiration for criteria, they can rely on the Examples-Diff feature to capture potentially missing details and ensure thoroughness. Currently, the Examples-Diff feature is optimized for binary classification sensing tasks, such as ``Is my desk cluttered?'' It generates meaningful ``positive'' and ``negative'' classes, like ``Cluttered desk'' vs. ``Uncluttered desk.'' Handling other types of questions and allowing users to more freely add, remove, or customize these categories can be addressed in future work.

\subsubsection{D4: Scaffold testing and debugging of criteria.}

Similar to best practices in software engineering, one important consideration when creating robust sensors is for users to test their sensors, ideally with edge cases that could lead to potential failures. Therefore, for each criterion, \systemname automatically generates two suggested test cases, with the flexibility to re-generate or produce more suggestions on demand (\fig\ref{fig:teaser}-g). Here, we again leverage the reasoning capabilities of foundation models, and instruct the model to first ``reason about how it might be challenging to assess a particular criterion based on different situations and scenarios that the user might encounter when using the sensor,'' and then generate ``test cases that are practical for users to try and test.'' This extends beyond common prompt engineering, which often focuses on adjusting the format and tone of the output based on a given input. \revision{Instead}, users are empowered to actively \revision{manipulate} their environment\revision{--for example, by} introducing new foreign objects or altering spatial configurations--in order to ``future proof'' their sensors. They can observe the model's responses and make targeted iterations on the criteria. The generated test suggestions are \revision{presented} as expandable chips for users to view and engage with.\looseness=-1

It is worth noting that, earlier in development, we \revision{explored generating broad}, top-level test suggestions based on the initial sensing task. \revision{While useful}, informal pilot testing revealed that these suggestions, were oftentimes too generic and \revision{lacked direct relevance} to users' specific criteria (e.g., testing under \revision{various} lighting conditions). Instead, we discovered that \revision{tailoring test suggestions to each individual criterion yielded} more actionable and effective results.
\revision{This approach sparked} deeper insights into the model's capabilities and limitations, facilitating users in refining their criteria \revision{more effectively and fostering a more robust testing process}.\looseness=-1

\subsection{Technical Implementation}

The \systemname web platform is developed using HTML, TypeScript, and CSS, utilizing the Lit Web Components library~\cite{lit_lit_2024} for building UI elements. A Python backend was implemented to manage LLM calls and additional API requests. All sensor data is stored locally using the browser's IndexedDB~\cite{bell_indexed_2023}.


Many of \systemname features are powered by the latest multimodal Gemini models \revision{as of October 2024}. Specifically, for running individual sensor criteria, we utilize the Gemini 1.5 Flash version~\cite{google_deepmind_gemini_2024} to ensure near-instant response times (see \fig\ref{fig:system}). We set the temperature to 0 to minimize output randomness. 
For tasks related to \revision{criteria creation} (e.g.,  automatically generating criteria and the examples-diff feature) \revision{as well as making the final verdict that aggregates results from individual criteria,} we \revision{employ} the Gemini 1.5 Pro version~\cite{google_deepmind_gemini_2024-1} for its advanced reasoning capabilities and \revision{support for long context window} (see \fig\ref{fig:system}). To encourage creativity in these tasks, we set the temperature to 0.8. 
Users have the flexibility to customize these default model configurations via a sensor's settings page.\looseness=-1 

However, it is important to note that our primary contributions lie more in the \textit{concept of scaffolded requirement elicitation and the design of user interface and experience for creating LLM-powered sensors}, which are independent of specific model usage. We anticipate that these designs will remain \revision{relevant} as generative AI models continue to advance in the near future.\looseness=-1

\section{User Study}\label{sec:eval}

\def\arraystretch{1.0}
\begin{table*}[t]
\vspace{-1mm}
\centering
\resizebox{1\textwidth}{!}{%
\begin{tabular}{p{42mm} | p{154mm}}
\toprule
\textbf{Metrics (Both Conditions)} & 
\textbf{Statement (7-point Likert scale)} 
\\\midrule
\textbf{Control} & 
With Prototype \{A, B\}, I felt I had control creating with the system. 
\\
\textbf{Understand Capabilities} & 
Prototype \{A, B\} helped me understand the underlying model’s capabilities – i.e. what it could and could not detect.
\\
\textbf{Communicate Requirements} & 
With Prototype \{A, B\}, I felt I was able to think through and communicate my personal requirements for the sensor.                       
\\
\textbf{Test} & 
With Prototype \{A, B\}, I was able to test and probe the sensor.                       
\\\midrule\midrule

\textbf{\systemname Metrics}  & 
\textbf{Statement (5-point Likert scale)}           
\\\midrule
\textbf{Manual Criteria} & 
How helpful was: The tool where I could make my own criteria toward helping you accomplish your goals?  
\\
\textbf{Automatic Criteria} & 
How helpful was: The tool that automatically generated criteria toward helping you accomplish your goals? 
\\
\textbf{Example-Diff} & 
How helpful was: The tool that generated criteria based on positive/negative examples (``Examples-Diff'') toward helping you accomplish your goals?
\\
\textbf{Multimodal Criteria} & 
How helpful was: The tool that let you provide images as examples for criteria toward helping you accomplish your goals? 
\\
\textbf{Test Cases} & 
How helpful was: The tool generated test cases for criteria toward helping you accomplish your goals? 
\\\bottomrule
\end{tabular}}
\caption{\textbf{Post-task questionnaire} filled out by participants after creating sensors. Participated rated both conditions for \textit{Control}, \textit{Understand Capabilities}, \textit{Communicate Requirements}, and \textit{Test} on a 7-point Likert scale. Then they rated the helpfulness of each of \systemname' features on a 5-point Likert scale.}
\Description{}
\label{tab:questionnaire}
\vspace{-6mm}
\end{table*}

To gather insights into \systemname' potential to benefit the AI sensor specification process, 
we conducted a 12-participant within-subjects user study. The study compares \systemname to a prompt-editor version, similar to the one used in the formative study (Figure \ref{fig:baseline-prompt-editor}), where participants specified AI sensor behavior by iterating on a text prompt. We also included \systemname' playback feature in the baseline condition, where participants could view the history of sensor outputs with their inputs.

\label{user-study-procedure}
\subsection{Procedure}
The overall outline of the study is as follows: (1) Prior to the study, participants completed a 30-minute self-directed tutorial, where they watched instructional videos and built a sensor with \systemname and the prompt-editor version. (2) During the study, participants spent 50 minutes creating two AI-powered sensors, starting either with \systemname (25 minutes) or the prompt-editor version (25 minutes), in a counterbalanced design. (3) After building these two sensors, participants completed a post-study questionnaire, which compared the two sensor prototyping conditions. (4) In a semi-structured interview, participants gave feedback on each prototyping tool, including the benefits and drawbacks of each one. The total time commitment of the study was 90 minutes.

From the brainstorm conducted in the formative study, we picked two different types of sensors for participants to implement in the user study, one ``urgent'' and one ``reminder'' sensor.
The two sensors were: (1) a ``reminder'' \textit{desk clutter} sensor which determines when the user’s desk is messy and (2) an ``urgent'' \textit{toddler safety} sensor which determines if there is anything particularly dangerous to a toddler in the user’s bedroom.
We chose these two sensors as they are realistic use cases of personal sensors, as well as general enough for users to be able to define and have personal opinions about them.
The order in which participants implemented these provided sensors was counterbalanced, in addition to the condition order.
To help situate the task, we asked participants to imagine that they were the target user for each sensor.
For the desktop clutter sensor, participants were asked to imagine that they were a professional who worked from home and was creating a sensor that would send a reminder for them to organize their desk during the work week if it got too messy.
For the toddler safety sensor, participants were asked to imagine that they were a new parent, building a sensor to help identify conditions in their living or bedroom which might be problematic for their toddler.\looseness=-1

Finally, participants authored and tested their sensors generally via their laptop camera, though some were able to use a webcam focused on the area of interest and author the sensor separately on their laptop. 
\revision{The study was approved by our
institution's IRB.}

\subsection{Participants}
We recruited 12 participants (6 female, 6 male) from our institution, covering a range of skill sets, including product managers, UX researchers, and software engineers, and from \revision{a variety of locations in the US, including Michigan, New York, California, Georgia, Wisconsin, and Washington}. 
We were generally recruiting for ``first-adopter'' tech-savvy individuals who would be likely candidates for authoring AI-powered sensors in their own home. 
Participants were recruited via an email invitation. 
The study was conducted remotely, in participants' homes to create a valid testing environment. 
Participants received a \$40 gift card for their participation.

\subsection{Questionnaire}
We wanted to understand the potential for \systemname to help participants think through their requirements for the sensor and steer the model to follow them. 
As such, our questionnaire (Table \ref{tab:questionnaire}) probes participants' self-perceived \textit{control} over the sensor, as well as their perceived ability to \textit{think through and communicate their personal requirements}. 
We were also interested in seeing how useful \systemname' \textit{test suggestions} were and if testing via modular criteria helped participants gauge the \textit{underlying model's capabilities}. 
We also included questions to assess the relative usefulness of \systemname' features.\looseness=-1

\revision{
\subsection{Data Analysis}
To analyze the results from the post-study questionnaire, we conducted paired sample Wilcoxon tests with full Bonferroni correction to compare the ratings from the two conditions, since the study was within subjects and the questionnaire collected ordinal data.

In addition, the main study sessions and the post-study interviews were screen and audio recorded, then transcribed. The first two authors independently coded the recordings and transcriptions using an open coding approach \cite{vaismoradi_content_2013} in accordance with Braun and Clarke's thematic analysis \cite{braun_using_2006}. Subsequently, they iteratively resolved disagreements and ambiguities, which included periodic discussions
with the research team. We present the key themes and findings below.
}

\section{Findings}\label{sec:findings}

\begin{figure}[t]
\centering
\includegraphics[width=.9\linewidth]{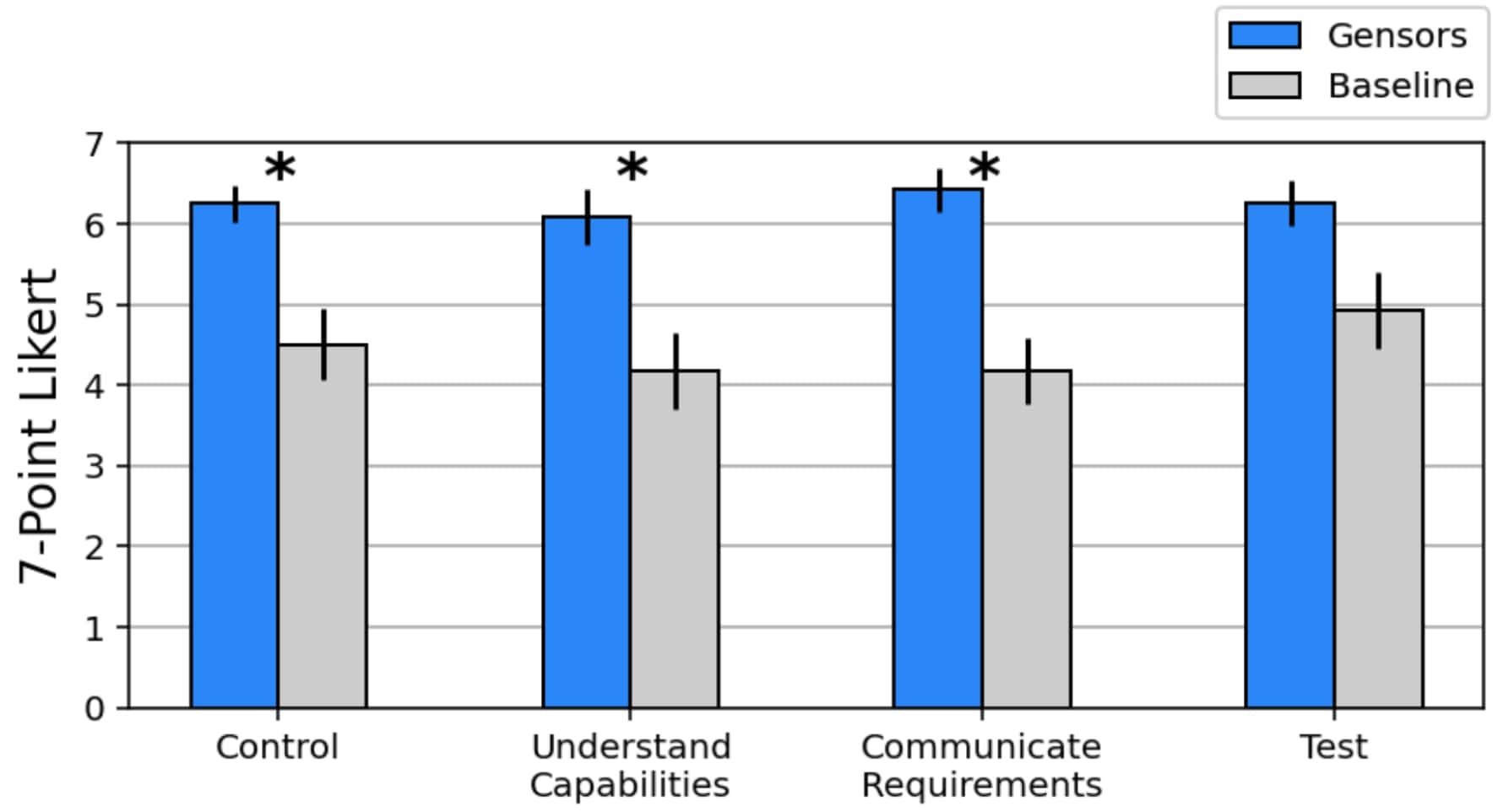}\hspace{5mm}
\vspace{-3mm}
\caption{\textbf{Comparing \systemname against baseline.} \systemname had significantly higher ratings for \textit{Control}, \textit{Understand Capabilities}, and \textit{Communicate Requirements}. (Bars are standard error and * indicates statistically significant difference, after full Bonferroni correction).}
\label{fig:stats-comparison-with-baseline}
\vspace{-3mm}
\end{figure}

\subsection{Quantitative Findings} 
\begin{figure}[t]
\centering
\includegraphics[width=.9\linewidth]{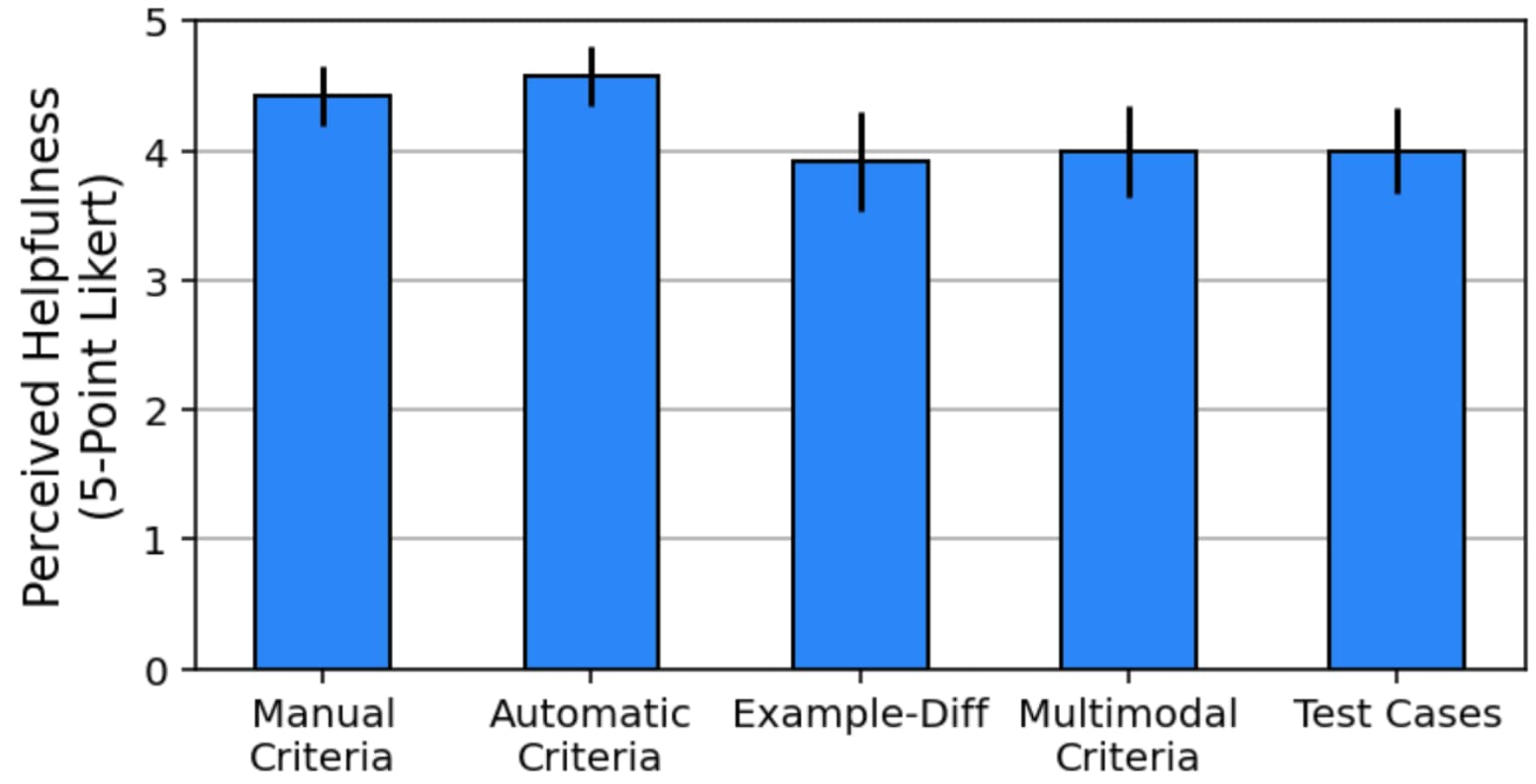}
\vspace{-3mm}
\caption{\textbf{Participants' feedback on \systemname features}. Of \systemname' features, all were considered helpful, with the automatically generated criteria perceived to be the most helpful.}
\label{fig:stats-system-features}
\vspace{-3mm}
\end{figure}

The results from the questionnaire are summarized in Figure \ref{fig:stats-comparison-with-baseline} \& \ref{fig:stats-system-features}. \revision{Notably,} participants \revision{reported} significantly \revision{greater} control over the sensor \revision{when using} \systemname ($\mu$ = 6.26, $\sigma$ = 0.72) \revision{compared to} the baseline ($\mu$ = 4.5, $\sigma$ = 1.44, p < .01). \revision{In the baseline condition,} participants were often \revision{uncertain about} how to best modify the prompt and felt their edits did little to influence the sensor. 

Participants' perceived understanding of the underlying model was significantly higher with \systemname ($\mu$ = 6.08, $\sigma$ = 1.11) than with the baseline ($\mu$ = 4.17, $\sigma$ = 1.57). With \systemname, participants could pinpoint which individual criteria the sensor was failing on, iterate on them, and understand whether that criteria could be \revision{indeed} detected by the model. 
\revision{In contrast, the baseline lacked the ability to} isolate different criteria, \revision{making it difficult for participants} to see how the model was attending to the requirements participants formulated in their prompts. 

Participants also rated their ability to think through and communicate their requirements with \systemname ($\mu$ = 6.42, $\sigma$ = 0.86) significantly higher than the baseline ($\mu$ = 4.17, $\sigma$ = 1.34, p < .001). This was predominantly \revision{attributed} to the automatically generated criteria (which provided a starting point for reflecting on requirements), as well as the Examples-diff feature (which helped participants find additional ``features'' to differentiate between borderline cases). 

Finally, \revision{although} the difference was not statistically significant, participants rated their ability to test their sensors higher with \systemname ($\mu$ = 6.25, $\sigma$ = 0.92) \revision{compared to the} baseline ($\mu$ = 4.95, $\sigma$ = 1.55).\looseness=-1

The two highest rated features of \systemname (see Figure \ref{fig:stats-system-features}) were \revision{the ability to} (1) automatically generate criteria ($\mu$ = 4.58, $\sigma$ = 0.76) and (2) manually create criteria ($\mu$ = 4.42, $\sigma$ = 0.76). The ability to define and debug the sensor's performance via individual criteria greatly supported users' workflows for creating sensors.

\subsection{How participants iterated on their sensors}
The two conditions led to markedly different workflows for iterating on sensors. With the baseline prototype, participants \revision{typically} observed the sensor's live output and spontaneously \revision{appended} more criteria and caveats to their prompt in an attempt to better align the \revision{model's behavior} with their own definition of the problem. 
For example, P1 \revision{began with a} simple prompt: ``\promptText{Is my desk cluttered? Output yes or no and explain why}.'' \revision{The model initially} stated that her desk was cluttered with too many items (e.g. a phone, pens, and keyboard). \revision{However,} P1 considered these items as permanent fixtures of her desk, so she updated the prompt with: ``\promptText{I do not consider it cluttered if my phone, keyboard, mouse, pens, and picture frame are on the desk}.'' 
Later, she specified that empty food and drink containers should be considered cluttered. \revision{However,} the model interpreted this too rigidly, \revision{flagging even a single container a clutter}. To fix this issue, she appended the clause: ``\promptText{One glass is not a problem}.''
\revision{Yet,} not all misinterpretations were easy to \revision{correct}. \revision{For instance,} P1 tried to define a \revision{specific} region of her desk to pay most attention to for clutter--`\promptText{between here and the keyboard}''--but the model could not grasp this distinction. After \revision{repeated attempts to rephrase, P1 expressed frustration, feeling feeling that her efforts were counterproductive:} \userquote{I've gone too far down this rabbit hole, and it's making it worse not better.} Ultimately, in the baseline condition, participants \textit{reactively} appended criteria and adjusted their prompts as they tested, spending a significant time tweaking their phrasing, as opposed to focusing predominantly on their criteria.\looseness=-1

\revision{In contrast}, with \systemname, participants \revision{adopted} a more deliberate and top-down \revision{workflow}, where they first defined a semi-comprehensive set of criteria and then revised these criteria through experimentation. 
For example, P2 started her desk clutter sensor with a few manually-written criteria, such as: ``\promptText{is the majority of the space in the view covered with objects?}'' 
When she was out of ideas, she decided to \revision{utilized the system's auto-generate feature to expand her criteria set}. The resulting suggestions, such as ``\promptText{are there non-work items on the desk?}'' complemented her initial set and created a more comprehensive set of requirements.
With these criteria \revision{established upfront}, participants then focused on refining them systematically \revision{rather than} opportunistically discovering and iterating on criteria as they came up, \revision{as seen in the baseline condition}. 
P6 specifically noted that he was \userquote{trying to debug each criteria one-by-one,} making adjustments to each one. 
Overall, \systemname encouraged a systematic, top-down workflow where participants first focused on the set of criteria that would control their sensor and then refined each of them individually. We \revision{further explore} how users debugged and revised their criteria in the following section.\looseness=-1

\subsection{\systemname helped participants tune and troubleshoot sensors via criteria-level controls}

A key benefit of \systemname was that participants could operate on their sensor at the criterion level, as opposed to the prompt level. Having the sensor explicitly check for each criteria helped participants better inspect and assess the model's \textbf{underlying reasoning}. In the following section, we discuss how employing the model's reasoning capabilities at the criteria-level helped participants with controlling, testing, and debugging their sensors.

\subsubsection{Participants felt they had greater control via criteria than with the prompt.}

When working on their prompts in the baseline condition, participants were often unsure of how to best improve their prompt and if it was improving \revision{at all}. 
\revision{For instance,} P5, working on the desktop-clutter sensor in this condition, \revision{aimed to have} the model output that his desk was cluttered when at least one empty can \revision{was present}. However, despite \revision{multiple adjustments} to the prompt, the sensor \revision{persistently reported} his desk as clean. He carefully scanned the prompt for any \revision{potential miscommunications} and hypothesized that the use of ``\promptText{and}'' in the phrase ``\promptText{Clutter also includes clothing accessories like hats \textbf{and} food waste like soda cans}'' might narrowly defined clutter to \revision{require both} both clothing accessories and food waste \revision{to be present}. So, he changed this ``\promptText{and}'' to ``\promptText{or}'', which then caused the sensor to briefly classify the desk as cluttered, \revision{though} not consistently so. \revision{Reflecting} on this, P5 stated that prompts are \userquote{easy to adjust, but it [the sensor] is not picking up the adjustments. I have to debug if it's my wording or the model misfiring.} 
When participants could not seem to influence the model, they began to question the prompt's overall structure and efficacy. For example, when P1 could not get her prompt to focus on a particular portion of her desk, she eventually deleted her current version and started \revision{over}. Overall, \revision{this sense of inadequacy in influencing the model often led to frustration, with some participants opting to restart the whole process in severe cases.}

Meanwhile, in the \systemname' condition, participants felt they could better influence the sensor's performance \revision{through their edits to the criteria}. All participants appreciated that each new criteria they added was explicitly checked, with an accompanying explanation. When the final verdict of the sensor wasn't exactly what they expected, they were generally able to \revision{identify} the corresponding criterion (or criteria) that wasn't being interpreted correctly and modify it. For example, P3's desk clutter sensor mistakenly \revision{classified} his desk as uncluttered \revision{despite} a stack of differently-sized books placed \revision{in the center}. He expected the criterion which checked if the desk felt ``\promptText{chaotic and disorganized}'' to be triggered by this stack of books; however, the system considered the stack to be organized. He was \revision{the} able to refine the criterion's wording to specify ``\promptText{book stacks with dissimilar-sized sized books}'' as chaotic, as well as add an image example, which ultimately steered his sensor to the results he \revision{expected}. With \systemname, participants could make targeted edits and could add targeted examples to criteria to steer the sensor, \revision{as summarized by P3}: \userquote{It's easier to be confident. It's easier to change one criterion than to rewrite the whole thing [prompt].}

\subsubsection{Participants could robustly test and debug with \systemname. 
}

In the baseline condition, \revision{participants faced challenges isolating specific criteria and testing them effectively. To overcome this,} they resorted to \revision{alternative} approaches, such as adjusting their prompt to \revision{elicit more} verbose descriptions of the scene. 
\revision{For instance,} while working on the toddler-safety sensor, P7 quickly added to her prompt: ``\promptText{Please list any hazards visible}.'' \revision{Similarly}, P4 added: ``\promptText{List all the objects you see and output if they are harmful to a toddler or not}.'' 
\revision{Despite} these detailed descriptions, isolating and testing criteria \revision{still proved quite difficult}. For instance, P4 was testing a clause in his prompt 
\revision{designed to check} for ``\promptText{precariously placed}'' items that might fall on the toddler. He \revision{experimented with} modifying the environment (e.g. placing a board on the edge of their bed), as well as adjusting the prompt to produce longer explanations. \revision{However}, the model consistently classified the room as safe without \revision{explicitly addressing} the ``\promptText{precariously placed}'' criterion. \revision{Despite further efforts to elicit longer explanations}, P4 \revision{ultimately remained uncertain about the model's reasoning regarding this particular criterion.} 
In summary, participants \revision{sought to} debug the sensor by requesting verbose explanations, but often struggled to get targeted \revision{insights} pertaining to a specific criterion.\looseness=-1

Meanwhile, with \systemname, participants could individually debug criteria, \revision{toggling} each one on and off, inspecting each criterion’s output one by one, and iterating on each one \revision{if needed}. For example, for the desktop clutter sensor, P6 first started with the  ``\promptText{object count}'' criterion \revision{while disabling the rest}. \revision{Observing} that this criterion was counting upwards of eight objects on their desk but still did not consider that enough for clutter, P6 \revision{refined it by adding} ``\promptText{More than 4 objects is clutter}'' to \revision{this specific criterion}. 
\revision{Next}, he moved on to the ``\promptText{non-work item}'' criterion, which checked for non-work items on the desk. He noticed that the sensor marked this criterion as ``true,'' stating in its explanation that there was a mug and flower on the desk. P6 thus adjusted this criterion to consider their mug and flower vase as permanent fixtures of the desk, not an indicator of messiness. 
\revision{With these refinements, P6 was satisfied with the performance of the two criteria,} enabled both, and quickly assessed the \revision{the sensor's overall effectiveness}. 
Through this isolated testing, participants were better able to understand the underlying model’s capabilities and \revision{integrate} the \revision{most effective} criteria \revision{into} their sensors. On rare occasions, some participants also removed criteria that did not seem to be working well \revision{deemed unnecessary}.\looseness=-1

\subsection{\systemname helped participants consider failure modes beyond their specific context}

In both conditions, participants manipulated their environment to test how their sensors would perform in a variety of scenarios. A common strategy was to create exaggerated setups, for example, \revision{making a desk conspicuously cluttered or impeccably clean}. After establishing that the sensor could work effectively in these \revision{clear-cut} scenarios, they would make \revision{incremental} adjustments, generally by adding or removing objects, to bring the scenario closer to a \textit{boundary} condition. 
For example, with \systemname, after his sensor stated his tidy kitchen was safe for toddlers, P9 placed a cutting knife on the countertop to \revision{test} the model's \revision{response to} a \revision{highly} unsafe object in an otherwise safe kitchen. 
Similarly, in the baseline condition, P5 \revision{organized} his desk to what he considered a clean state, then introduced individual perturbations to make it ``messy,'' such as adding a can of soda. Overall, participants manipulated their environments to explore and test their sensors \revision{across} a variety of scenarios.\looseness=-1

While participants naturally engaged in creating these tests, they \revision{tended to overlook} tricky visual situations or subtler failure modes in the baseline condition. Meanwhile, some participants appreciated that \systemname created suggested test cases for them, noting that it would otherwise be difficult to think of these alternative scenarios on their own. 
\revision{For example, P10, working on the desk clutter sensor with an ``\promptText{object count}'' criterion}, appreciated the test case that suggested placing similar colored objects stacked on their desk to see if the model could distinguish them. They found that stacking objects together did impede the sensor's ability to clearly distinguish them, especially if parts of objects were hidden beneath others. 
Similarly, for the ``\promptText{choking hazard}'' criterion in his toddler safety sensor, P4 appreciated a test case that suggested partially obscuring the choking hazard object (e.g. a coin) with another larger object. \revision{Overall, these criterion-specific test suggestions helped participants more thoroughly stress-test their sensors while gaining a deeper understanding of the underlying MLLM's capabilities.}



\subsection{\systemname~ \revision{supplemented} users' own ``blind spots'' by suggesting complementary criteria}

While \systemname~ \revision{enabled} users \revision{to align} the sensor \revision{most closely with} their own personal criteria and preferences (e.g. via the manual-criteria feature), it also \revision{encouraged them to look beyond their immediate surroundings} and field of view (e.g. via the auto-generated criteria and Examples-diff features) \revision{for criteria that they would otherwise overlook}. 

For example, some users noted that the automatically-generated criteria \revision{prompted} them \revision{to consider aspects of the problem} they would not have thought of \revision{independently}. For example, many users did not initially think of ``\promptText{uncovered outlets}'' as a potential hazard for toddlers, but realized they had missed this upon seeing it auto-generated. 
Similarly, for the desk clutter sensor, \systemname considered whether there was a ``\promptText{Usable Area}'' on the desk (e.g. ``\promptText{Is it easy to find a usable area on the desk without having to move items around?}''). \revision{Participants found this surprising and compelling, as it reframed the problem of messiness in an unexpected way.} 
\revision{However, one participant, P1}, \revision{commented on a potential drawback of generated criteria, noting that they seemed to be} \userquote{too easy to say yes to,} which might distract users from critically thinking about their own preferences. He then imagined alternative workflows where \systemname could help users actively consider criteria suggestions by posing yes/no questions.

Furthermore, the Examples-diff feature \revision{proved instrumental in} helping users overcome ``tunnel vision'' when they \revision{struggled to identify additional} criteria that could further differentiate borderline cases. 
\revision{For instance}, while P11 was working on their desktop clutter sensor, they had a few borderline examples that their current set of criteria was not consistently differentiating. At this point, they felt that they had exhausted the criteria they could identify, so they \revision{turned to} the Examples-diff feature, from which they selected two criteria: one that checks if the ``\promptText{objects on the desk appear to be randomly placed}'' and another that checks if the desk feels ``\promptText{chaotic and disorganized}.'' 
With these criteria, P11 felt his borderline examples were more consistently distinguished: \userquote{It's easy to describe the easy cases [with criteria]. It's hard to describe the boundary cases. Having the model say here's the difference between these two lets you focus on what the model is seeing.} 
\revision{Occasionally}, however, the Examples-diff feature would hallucinate differences between examples. \revision{For instance}, P12 used the feature for her toddler safety sensor, and the model hallucinated that there was a toilet in their bedroom, creating a criterion that checked if the toilet's lid was closed. 
\revision{Despite these occasional inaccuracies}, the Examples-diff feature generally helped participants \revision{uncover} more features to distinguish between more complex examples.\looseness=-1

\subsection{How MLLM peculiarities impacted sensor creation}

MLLM hallucinations both benefited and inhibited participants' processes in defining their sensor criteria. \revision{On one hand, when not entirely fabricated, hallucinations could sometimes be useful for helping participants become aware of criteria they did not consider.} 
For example, while P3 was building his toddler safety sensor in baseline, the model deemed his bedroom unsafe \revision{due to the potential risk of a toddler falling out of open windows}. \revision{Although P3's windows were not actually open, he found this scenario plausible and ultimately adjusted his prompt to account for such a situation.}\looseness=-1 

\revision{On the other hand}, entirely fabricated hallucinations could confuse and distract users \revision{during the process}. 
\revision{For instance}, while P10 was working on her desk clutter sensor in the baseline, the model \revision{falsely reported the presence of a stuffed animal on her desk}. Hoping to prevent this hallucination in future runs, P10 added the clause ``\promptText{Ps: I don't have stuffed animals}'' to her prompt.
\revision{As a potential workaround}, both P6 and P10 wanted a visual explanation (e.g. a bounding box around the supposed ``stuffed animal'') to better understand \revision{the model's perception}. 

In both conditions, the \revision{stochastic nature} of MLLMs occasionally \revision{produced} different results across runs with \revision{nearly identical} inputs, which both confused and informed participants. 
Like traditional ML sensors, participants' MLLM-powered sensors decisions \revision{exhibited ``flickering'' behavior}, for example, \revision{alternating between} ``cluttered'' and ``uncluttered'' assessments for the same desk setup. 
But perhaps what was more disorienting was that the accompanying explanation also \revision{varied} across each run. 
Interestingly, P1, P2, and P12 all interpreted flickering as the sensor being ``unsure'' and used flickering as a \revision{cue} that they needed to refine their prompt or criteria. 
In the baseline condition, P1 noticed her desk clutter sensor was flickering, \revision{and hypothesized that it was} due to a clause in her prompt that specified ``\promptText{lots of papers}'' as clutter; she could see the model's explanation \revision{alternated} between \revision{claiming that there were} ``too many papers'' to ``just a few.'' 
P12 gleaned a bit more information from flickering by aggregating the model's decisions over a window (e.g. the sensor output ``cluttered'' the past 3 of 5 runs) to \revision{gauge} which way the model was leaning towards and to better inform debugging. 
Ultimately,the stochasticity of the underlying model sometimes \revision{became a tool} for participants to assess its uncertainty and the current state of their criteria.\looseness=-1

\section{Discussion}

\subsection{Supporting Active Criteria Elicitation}
While participants appreciated the automatically generated criteria, some noted that it could potentially stifle them from carefully considering their own personal preferences for the sensors. \revision{They desired features that would more actively involve them in creating the criteria.}
There are many possibilities for alternative, more \revision{engaging} workflows to help users realize their requirements. 
For instance, for a desk clutter sensor, one could first be presented with a set of personas, such as an ``organized chaos'' persona, \revision{who considers an uncluttered desk to be one that, while containing many items,is arranged in a neat and meticulously organized manner}, or \revision{a ``minimalist'', who prefers a desk with very few items.}
\revision{Users can reflect on these personas and their own preferences to \textit{select the one they most identify with, which would then determine their initial set of criteria.}}
Another, perhaps simpler mechanism to actively engage users in thinking through their personal criteria \revision{could involve} posing yes/no questions (e.g. ``Do you typically keep a lot of items on your desk?''). Future research \revision{could explore the potential advantages and limitations of these alternative workflows for criteria elicitation.}\looseness=-1

\subsection{An IoT Network of Sensors Proactively Reasoning Together}
Reasoning can be extended beyond helping users \revision{achieve a particular sensing objective} to creating an Internet of Things (i.e. a network of sensors) that communicate with each other and \textit{proactively reason}. Instead of \revision{relying on a central node for decision-making}, each node might reason over its own data and make proactive requests of other nodes, such as requesting data or suggesting actions. 
\revision{For instance, a user's smartwatch might detect a marathon training pattern from analyzing recent running data. It could then proactively request information from the user's smart fridge to check if the necessary foods are available to support this training regimen. The fridge would analyze its contents, reason about missing items, and inform the smartwatch, which could then prompt the user to go grocery shopping.}\looseness=-1

\revision{Beyond this initial interaction, the fridge could proactively reason about further analyses it could conduct to continue supporting the user, such as by tracking their eating habits over the coming weeks (e.g., it might assess whether the user is consuming enough protein or carbohydrates for marathon preparation).} To fulfill its analysis, the fridge might request fitness data from the smartwatch, such as the user's weight and recent runs, to offer detailed meal recommendations. Overall, there are exciting possibilities for sensors within IoT networks to \revision{independently reason with their data, anticipate user needs, and collaborate with other sensors to provide proactive support.}\looseness=-1

\subsection{Denoising Sensors via Reasoning}
Another \revision{promising} application of reasoning is in \textit{denoising} AI sensors.
Currently, temporary occlusions or disturbances in a sensor's data stream can lead to erroneous decisions.
\revision{For example, consider a sensor designed to track the duration of a user's painting practice. At the start of a session, a pet might wander in front of the sensor’s camera, temporarily obscuring the user. This could cause the sensor to mistakenly conclude that the user has stopped painting.} To address such noise, the model could incorporate reasoning about the \textit{task state} and \textit{recent outputs}. For instance, the model might infer that (1) painting is generally a longer task, and (2) that the user had just started painting based on prior video frames. Even with the temporary occlusion caused by the pet, the sensor could reasonably assume that the user is still painting.
\revision{Future research could further explore equipping sensors with reasoning capabilities to enhance their resilience against noise by integrating contextual understanding of tasks and leveraging prior data outputs.}

\section{Limitations \& Future Work}

This work represents a first step towards user-friendly AI sensor definition with MLLMs, though several limitations suggest areas for future research.

\revision{We acknowledge that our study sample size of 12 participants, all relatively tech-savvy, may impact the generalizability of our findings. While we sought diversity in participant backgrounds and skill sets, future research could involve a larger, less tech-oriented sample to provide a more balanced interpretation of how users engage with \systemname. This would help ensure broader applicability of the results.}
In addition, we observed that users sometimes tested their sensors with limited example diversity, raising concerns about generalizability. Future iterations of \systemname could incorporate mechanisms to incentivize broader testing and evaluation across diverse scenarios, ensuring robust performance across a wider range of situations. 
Similarly, our study focused on a limited set of use cases, and some participants expressed less familiarity with scenarios outside their personal experience, such as those related to parenting. \revision{Expanding the range of use cases and allowing participants to prioritize scenarios they are familiar with could provide further insights.}\looseness=-1

To gain a more comprehensive understanding of user needs and challenges, future research should move beyond the controlled lab setting and explore longitudinal studies where participants deploy and interact with AI sensors in their natural environment. This would provide valuable insights into long-term usability and identify challenges related to real-world deployment, such as integration with other smart home devices, enabling more complex actions triggered by sensor outputs, and exploring the potential for collaborative sensor development and sharing. This could also address limitations related to camera placement, field of view, and image quality, which were observed to impact sensor performance. Future work could explore providing standardized camera setups, developing tools for automatic scene adjustment, or incorporating depth information to enhance scene understanding.

In addition, addressing inherent limitations of MLLMs, such as hallucinations, is crucial. These hallucinations occasionally confused users and hindered their ability to understand and verify a sensor's behavior. \revision{On one hand,} developing techniques to improve explainability and transparency, such as visualizing the model's focus on an image frame through bounding boxes or highlighting \cite{song_popup_2019}, could be beneficial. \revision{On the other hand, as the cost and latency associated with model calls continue to decrease, it may become feasible to run the same sensor prompt under varying temperatures or configurations and implement a subsequent ``majority voting'' mechanism. This approach, akin to how a final verdict prompt aggregates and determines an outcome based on individual criteria, offers promise in addressing the stochastic nature of MLLMs and mitigating hallucinations.
}\looseness=-1

\revision{Furthermore,} enhancing \systemname to support richer logical constructs beyond basic AND/OR operators and \revision{to handle continuous values or fuzzy boundaries for sensor outputs will broaden its applications and better capture real-world nuances.} 
This may involve integrating confidence levels, thresholds, or fuzzy logic to \revision{deliver more informative and flexible sensor outputs. 
However, it will be crucial to balance this increased complexity with the implications on usability.}\looseness=-1

\revision{Last but not least, leveraging cloud-based LLMs could raise potential privacy concerns, particularly the risk of exposing sensitive home images or audio upon security breaches. Additionally, users may have reservations about their data being used for further model training by the cloud provider. As a first step towards mitigating these risks in a research prototype, we use the Gemini API, which follows strict data retention policies to ensure no user data is stored or used for training. In the future, smaller and more capable multimodal models could enable more inference tasks to be processed directly on users' devices, thereby significantly reducing security and privacy risks by limiting data transmission to external servers. This shift toward on-device intelligence offers a promising path for enhancing the privacy and security of personalized sensing systems.
}

\section{Conclusion}

This work explored using MLLMs to create customizable AI sensors, where users define complex sensing tasks through natural language. However, our formative study revealed challenges in effectively articulating requirements and specifying desired sensor behavior through basic prompting.  To address this, we developed \systemname, a system that facilitates AI sensor definition by decomposing high-level sensing tasks into explicit, testable criteria. \systemname leverages MLLM capabilities to offer relevant criteria, enable user customization, translate examples into new criteria, and suggest tests. Our user study demonstrated that \systemname significantly enhanced the AI sensor definition process, fostering deeper understanding, more systematic exploration of model capabilities, and ultimately, more robust and personalized sensor definitions.
\revision{Despite challenges such as model hallucinations,} \systemname demonstrates the potential to make intelligent sensing technologies more accessible and customizable. Future work could focus on mitigating these limitations and further enhancing user experience and control over sensor behavior.\looseness=-1

\begin{acks}
We gratefully thank Jonathan Caton, Adam Connors, Aaron Donsbach, Lucas Dixon, Noah Fiedel, Jeff Gray, Minsuk Kahng, Alejandra Molina, Crystal Qian, Fernanda Viegas, Wing Wat, Martin Wattenberg, and Xiaotong Yang
for their helpful comments. 
We also appreciate the valuable input from
our study participants.
\end{acks}

\bibliographystyle{ACM-Reference-Format}
\bibliography{references}


\end{document}